\documentclass{PoS}

\title{Recent Results from Daya Bay}

\ShortTitle{Recent Results from Daya Bay}

\author{\speaker{Liang Zhan}\thanks{E-mail: zhanl@ihep.ac.cn}\\
        Institute of High Energy Physics, Beijing\\
        On behalf of the Daya Bay Collaboration}

\abstract{
The Daya Bay reactor neutrino experiment observed electron antineutrino disappearance associated with $\theta_{13}$ with a significance better than $5\sigma$ in 2012. The final two of eight antineutrino detectors were installed in the summer of 2012. Including the 404 days of data collected with the full detector configuration resulted in a 3.6 times increase of statistics over the previous result with the 6-AD configuration. With improvements of the systematic uncertainties and better estimation of backgrounds, Daya Bay has measured $\sin^22\theta_{13} = 0.084\pm0.005$ and $|\Delta m^2_{ee}|=2.42^{+0.10}_{-0.11}\times 10^{-3}$~eV$^2$. This is the most precise measurement of $\sin^22\theta_{13}$ to date and the most precise measurement of of $|\Delta m^2_{ee}|$ via electron antineutrino disappearance. Several other analysis results are presented, including an independent measurement of $\theta_{13}$ using inverse-beta decays associated with neutron capture on hydrogen, a measurement of reactor antineutrino flux and spectrum, and a search for light sterile neutrino mixing.}

\FullConference{XVI International Workshop on Neutrino Telescopes,\\
		2-6 March 2015\\
		Palazzo Franchetti – Istituto Veneto, Venice, Italy }

\makeatletter
\def\@oddfoot{\ifnum\thepage=1%
  \PoScopyright@box\hfill%
  \if@PoSspecialurl\PoSspecial@url\else\unhbox\PoSpaper@url\fi\fi}
\makeatother

\begin{document}

\section{Introduction}
The observations of neutrino oscillation from the sun, the atmosphere, reactors, and particle beams have established that neutrinos have mass and that three flavor states $(\nu_e, \nu_{\mu}, \nu_{\tau})$ are superpositions of three mass states $(\nu_1, \nu_{2}, \nu_{2})$. The mixing can be quantified using a $3\times 3$ matrix, known as the Pontecorvo-Maki-Nakagawa-Sakata (PMNS) mixing matrix, which can be parameterized in terms of three mixing angles $(\theta_{12}, \theta_{23}, \theta_{13})$ and a CP violating phase ($\delta$). The neutrino oscillation also depends on two mass-squared differences $\Delta m^2_{21}$ and $\Delta m^2_{31}$.

Among the three mixing angles, the value of $\theta_{13}$ can be extracted using the electron antineutrinos from the reactors at a short baseline ($\sim$km). The survival probability is given by
\begin{equation}
\label{eq:psurv}
P_{\overline{\nu}_e\rightarrow\overline{\nu}_e} = 1
-\cos^4\theta_{13}\sin^2 2\theta_{12}\sin^2\Delta_{21}
-\sin^2 2\theta_{13}(\cos^2\theta_{12}\sin^2\Delta_{31} + \sin^2\theta_{12}\sin^2{\Delta_{32}}),
\end{equation}
\noindent where $\Delta_{ji}\equiv 1.267 {\Delta}m^2_{ji}({\rm eV}^2)\frac{L({\rm m})}{E({\rm MeV})}$, and ${\Delta}m^2_{ji}$ is the difference
between the mass-squares of the mass eigenstates $\nu_j$ and $\nu_i$.
Besides the mixing angle $\theta_{13}$, the effective mass-squared difference, $\Delta m^2_{ee}$, defined by $\sin^2\Delta_{ee} \equiv \cos^2\theta_{12}\sin^2
\Delta_{31}+\sin^2\theta_{12}\sin^2{\Delta_{32}}$, can also be
determined by measurement of the energy-dependent oscillation.

In 2012, the Daya Bay experiment observed a non zero $\theta_{13}$ at a significance of 5.2$\sigma$~\cite{DYB1} via a rate-only measurement with 55 days of data. Other experiments observed consistent results~\cite{T2K, MINOS, DC1, RENO}. In 2013, Daya Bay updated the results with 217 days of data, with a spectral analysis to measure the oscillation frequency, which led to the first direct measurement of the effective mass-squared difference $\Delta m^2_{ee}$ in the electron antineutrino disappearance channel. The relative large value of $\theta_{13}$ and its precise measurement facilitate future neutrino oscillation research on the neutrino mass hierarchy and CP violating phase.

\section{The Daya Bay Experiment}
The Daya Bay Reactor Neutrino Experiment was designed to make a precise measurement of the mixing angle $\theta_{13}$. The precision is ensured by the following features: 1) high statistical precision due to large thermal power of reactors and large target mass of detectors; 2) reduction of reactor related uncertainties by relative Far/Near measurements; 3) reduction of detector related uncertainties by identically designed multiple detectors; 4) low background by sufficient overburden and good shielding.

A detailed description of Daya Bay experiment can be found in Refs.~\cite{DYB_cdr, DYB_nima}. The Daya Bay experiment is located near three nuclear power plants (Daya Bay, Ling Ao, and Ling Ao II) on the southern coast of China. Each power plant has a pair of reactor cores with a distance of 88~m. All six cores are functionally identical pressurized water reactors, each with a maximum of 2.9 GW thermal power.
A total of 8 identically designed antineutrino detectors (ADs), with a 20-ton target mass each, are placed in three underground experimental halls (EH1 and EH2 for the near sites, and EH3 for the far site), covering baselines ranging from 360 m (EH1 to Daya Bay) to 1910 m (EH3 to Daya Bay).

Each AD consists of three nested cylindrical zones separated by two concentric acrylic vessels. The innermost zone contains 20 tons of gadolinium-doped liquid scintillator (Gd-LS). The middle zone is filled with 21 tons of un-doped liquid scintillator (LS), while the outer zone consists of 37 tons of mineral oil. The $\overline{\nu}_{e}$ is detected via the inverse $\beta$-decay (IBD) reaction, $\overline{\nu}_{e} + p \to e^{+} + n$, in the Gd-LS. The coincidence of the prompt positron light with the delayed gamma rays (totalling $\sim\!8$~MeV) generated from the neutron capture on Gd with a mean capture time of $\sim\!30~\mu{\rm s}$ provides a distinctive $\overline{\nu}_{e}$ signature. In each AD, light created as a result of
particle interactions in the Gd-LS and LS, is collected by 192 radially positioned 20-cm PMTs in the MO. The detectors have a light yield of $\sim$165 photoelectrons/MeV and a
reconstructed energy resolution of $\sigma_E/E\approx 8\%$ at 1 MeV. Three automated
calibration units (ACUs) mounted on the stainless steel vessel lid allow for remote deployment of a light-emitting diode (LEDs) and calibration sources, such as $^{68}$Ge, $^{241}$Am$^{13}$C and $^{60}$Co,  into the Gd-LS and LS volumes along three vertical axes.
The ADs in each EH are shielded with $>2.5$~m of high-purity water against ambient radiation in all directions. Each water pool is segmented into inner and outer water shields (IWS and OWS) and instrumented with photomultiplier tubes (PMTs) to function as
Cherenkov-radiation detectors whose data were used by offline software to remove spallation neutrons and other cosmogenic backgrounds. The detection efficiency for
long-track muons is $>99.7\%$~\cite{DYB_muon}. The water pool is covered with an array of resistive plate chambers (RPCs).

\begin{figure}[htb]
\centering
\includegraphics[width=0.6\columnwidth]{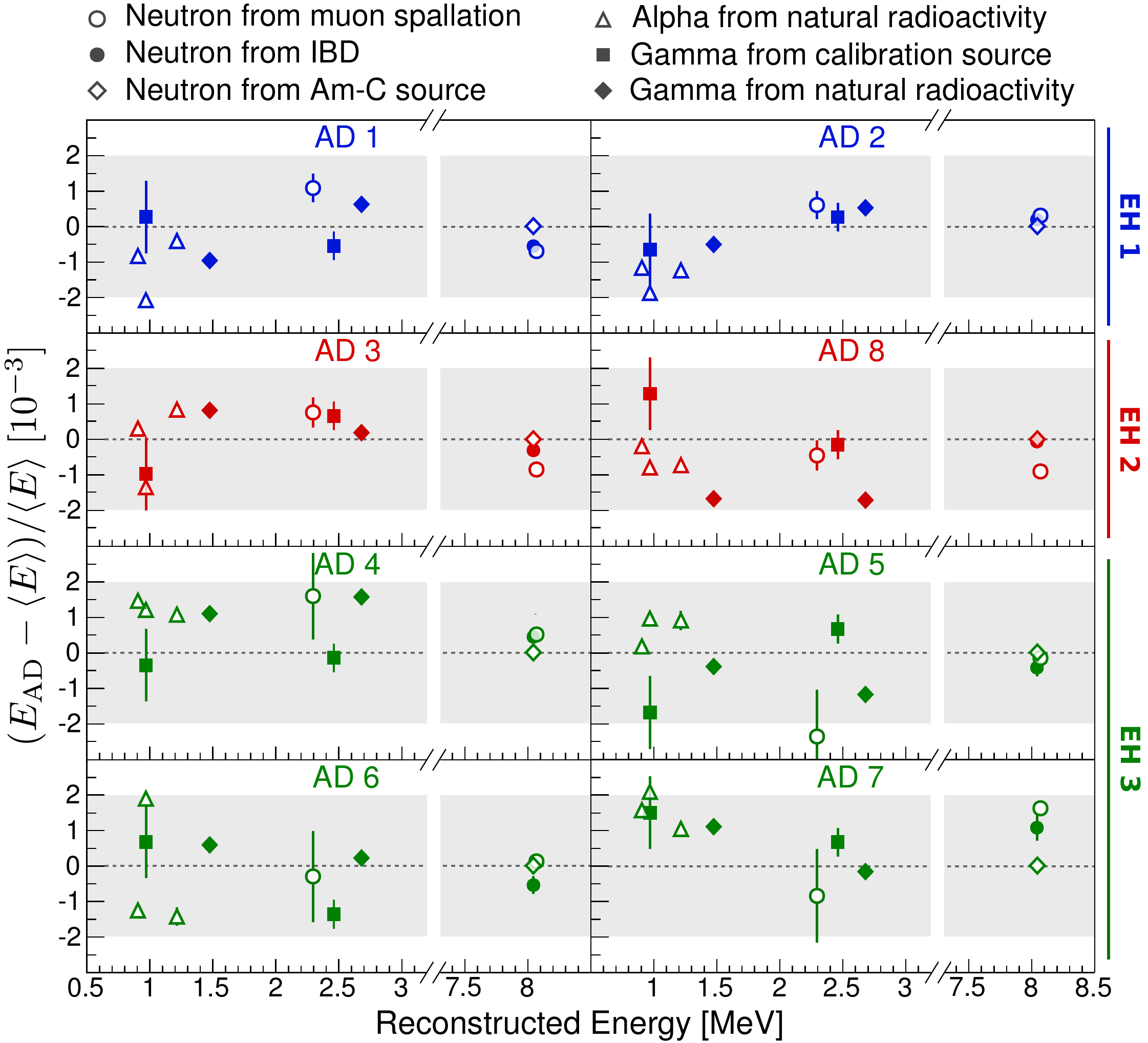}
\caption{
  Comparison of the reconstructed energy of calibration sources between antineutrino detectors.  $E_{\rm AD}$ is the reconstructed energy determined for each AD, and $\left\langle E \right\rangle$ is the 8-detector average.  Error bars are statistical only, and the difference between detectors for all calibration sources are $<$~0.2\%.
      \label{fig:escale}}
\end{figure}

The detector energy scale of all ADs was calibrated using $^{241}$Am$^{13}$C sources providing an $\sim$8~MeV peaks from neutron capture on Gd. The time variation and the position dependence of the energy scale was corrected using the 2.506~MeV gamma-ray peak from $^{60}$Co sources. The reconstructed energies of various calibration sources in different ADs are compared in Fig.~\ref{fig:escale}.
The ACU calibration sources $^{241}$Am$^{13}$C and $^{60}$Co were positioned at the detector center. Neutrons from IBD and muon spallation captured on Gd were nearly uniformly distributed in the Gd-LS region. $\alpha$ particles from polonium decays with positions estimated to be within the Gd-LS region provided another diffuse reference. Comparisons of intrinsic $^{40}$K, $^{208}$Tl, and spallation neutron capture on hydrogen were biased by backgrounds, therefore only those interactions with reconstructed positions within 1~m of the center of each detector were considered. Based on all this information, the uncorrelated uncertainty of the energy scale was determined to be 0.2\% in the energy range of reactor $\overline{\nu}_{e}$.

The energy response of the ADs is not linear due to scintillator and electronics effects, each contributing at the level of 10\%. The scintillator nonlinearity is particle and energy dependent, and is related to intrinsic scintillator quenching and Cherenkov light emission. The electronics nonlinearity is introduced due to the interaction of the scintillation light time profile and the charge collection of the front-end electronics.
The energy nonlinearity model was determined by a combined fit to monoenergetic $\gamma$ peaks from radioactive sources and the continuous $\gamma + \beta$ spectrum extracted from $^{12}$B produced by a spallation neutron after muon interactions with carbon. The electronics nonlinearity was determined by studying the time profile of charge in the data and MC.
The uncertainties of the nonlinearity are $<1\%$ above 2~MeV.

The Daya Bay experiment started data-taking on September 23, 2011 with two ADs installed in EH1. On December 24, 2011, the "6-AD" run started shortly after four more ADs were installed in EH2 and EH3. The previous results~\cite{DYB1, DYB_cpc, DYB_6AD_shape} on the measurements of oscillation parameters are based on the 6-AD data. The "8-AD" run started on October, 2012 after the installation of the final two ADs in the summer of 2012. The 8-AD data obtained up to Nov. 27, 2013 (corresponding to 404 days) were analysed together with the 6-AD data to extract the latest results of $\theta_{13}$ and $\Delta m^2_{ee}$~\cite{DYB_8AD}. An independent measurement of $\theta_{13}$ using the neutron capture on hydrogen of the 6-AD data has also been performed~\cite{DYB_nH}. The measurement of the reactor antineutrino flux and spectrum has been obtained for the 6-AD data~\cite{6AD_reactor}. Moreover, a search for light sterile neutrino using the 6-AD data has been reported~\cite{DYB_sterile}. In the next section, we will describe these recent results.

\section{Recent Results}
\subsection{Oscillation Analysis from Neutron Capture on Gadolinium}
A total of $\sim$1.1 million ($\sim$150 k) IBD candidates for the near (far) sites were selected using the same criteria described in~\cite{DYB1}. The IBD rate as well as the prompt and delayed energy spectra show consistency between side-by-side detectors. The relative energy scale uncertainty was improved to $0.2\%$. This reduction of 43\% compared to previous publication~\cite{DYB_6AD_shape} was achieved by improvements in the correction of position and time dependence, and enhanced the precision of $\Delta m^2_{ee}$ by 9\%.

Estimates for the five major sources of background for the new data sample are improved with respect to Ref.~\cite{DYB_6AD_shape}. Two of the three $^{241}$Am$^{13}$C sources in each AD in EH3 were removed during the 2012 summer installation period. As a result, the average correlated $^{241}$Am$^{13}$C background rate in the far hall decreased by a factor of 4 in the 8-AD period. The estimate of the fast neutron background was improved by tagging the fast neutron candidates following cosmogenic signals detected by the OWS or RPC. The energy spectrum of these veto-tagged signals was consistent with the spectrum of IBD-like candidate signals above 12~MeV, and was used to estimate the rate and energy spectrum for the fast neutron background in the range of 0.7-12~MeV. The methods in Refs.~\cite{DYB1, DYB_cpc} were used to estimate the backgrounds from the accidental background, the correlated $\beta-n$ decays from cosmogenic $^9$Li and $^8$He, and the $^{13}$C($\alpha$, n)$^{16}$O reaction for the current 6+8 AD data. The total backgrounds amounted to about 3\% (2\%) of the IBD candidates in the far (near) hall(s).

\begin{figure}[htb]
\centering
\includegraphics[width=0.48\columnwidth]{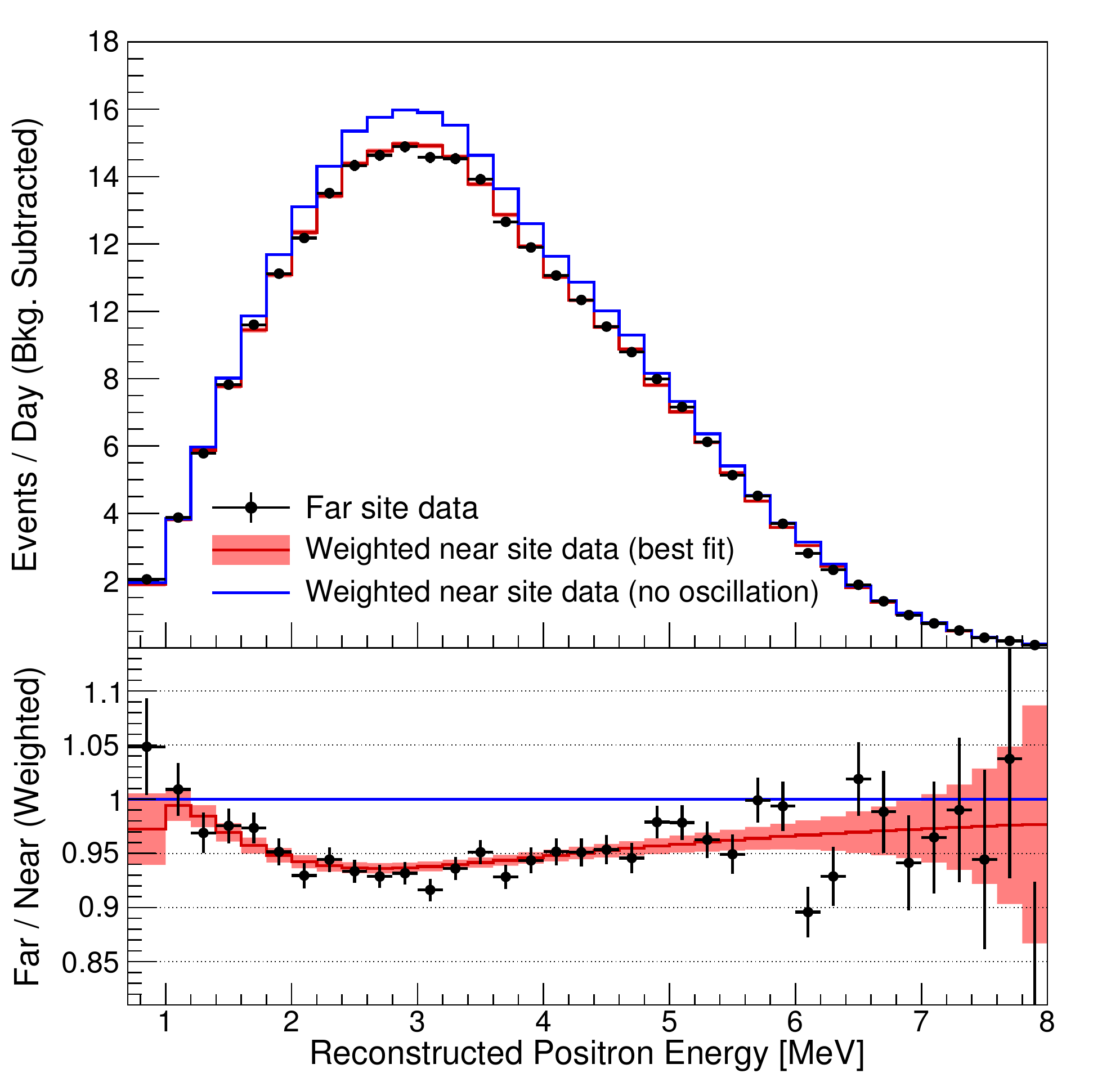}
\includegraphics[width=0.48\columnwidth]{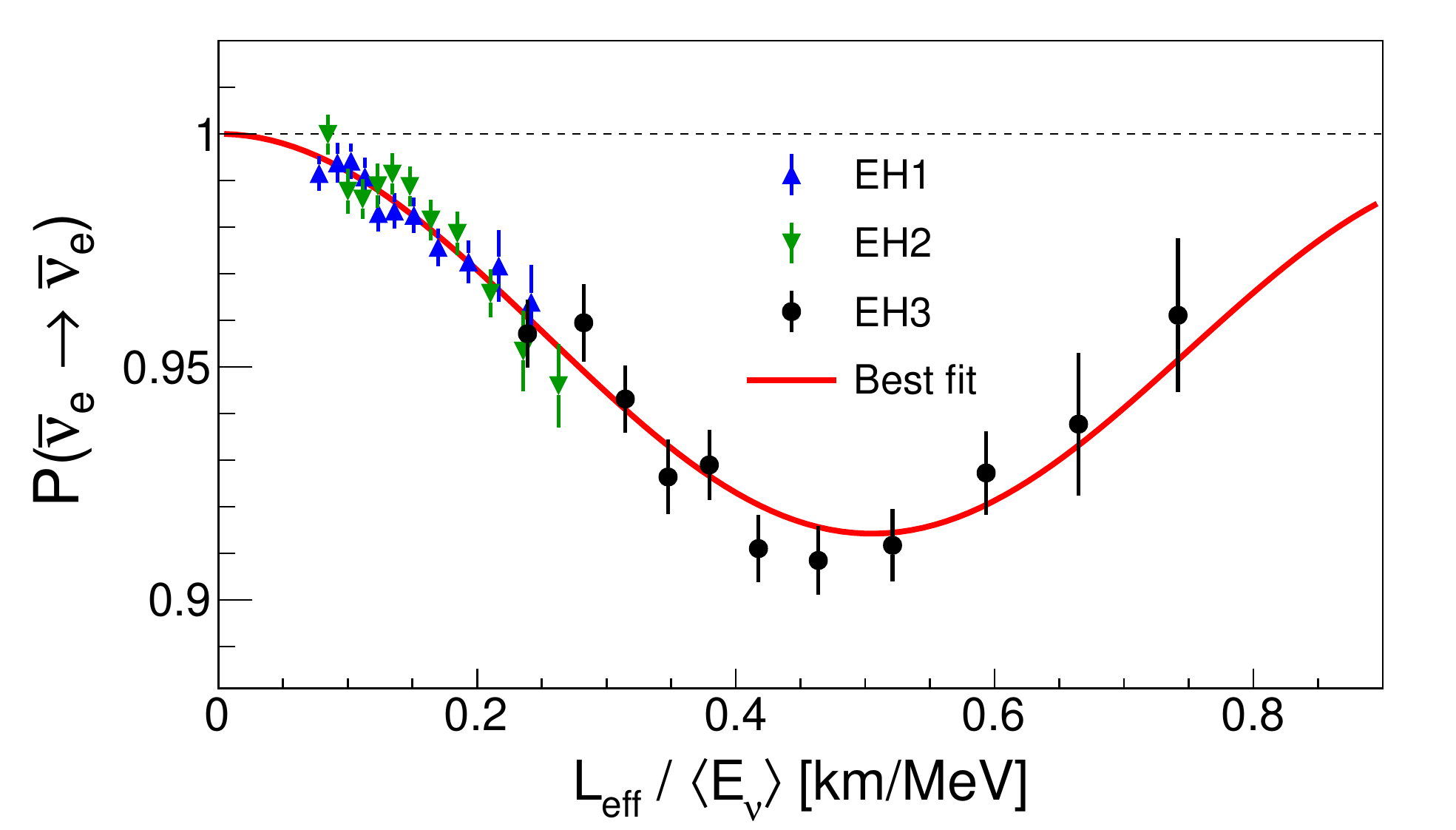}
\caption{Left: (Top panel) Background-subtracted reconstructed positron energy spectrum observed in the far site (black points), as well as the expectation derived from the
  near sites excluding (blue line) or including (red line) the best estimate of oscillation. (Bottom panel) Ratio of the spectra to the no-oscillation case. The error bars show the statistical uncertainty of the far site data. The shaded area includes the systematic and statistical uncertainties from the near site data.
  Right: Electron antineutrino survival probability versus effective propagation distance $L_{\mathrm{eff}}$ divided by the average antineutrino energy $\langle E_{\nu} \rangle$. The data points represent the ratios of the observed antineutrino spectra to the expectation assuming no oscillation. The solid line represents the expectation using the best estimates of $\sin^22\theta_{13}$ and $|\Delta m^2_{ee}|$. The error bars are statistical only.
\label{fig:spectracomp}}
\end{figure}

The reconstructed prompt energy spectrum observed in the far site is compared with the expectation based on the near-site measurements in left plot of Fig.~\ref{fig:spectracomp}. In the right plot of Fig.~\ref{fig:spectracomp}, the measured survival probability as a function of $L/E$ from all experimental halls is compared to the expectation assuming the best fit oscillation parameters. The spectrum distortion is highly consistent with the oscillation interpretation in the three neutrino framework.

\begin{figure}[htb]
\centering
\includegraphics[width=0.6\columnwidth]{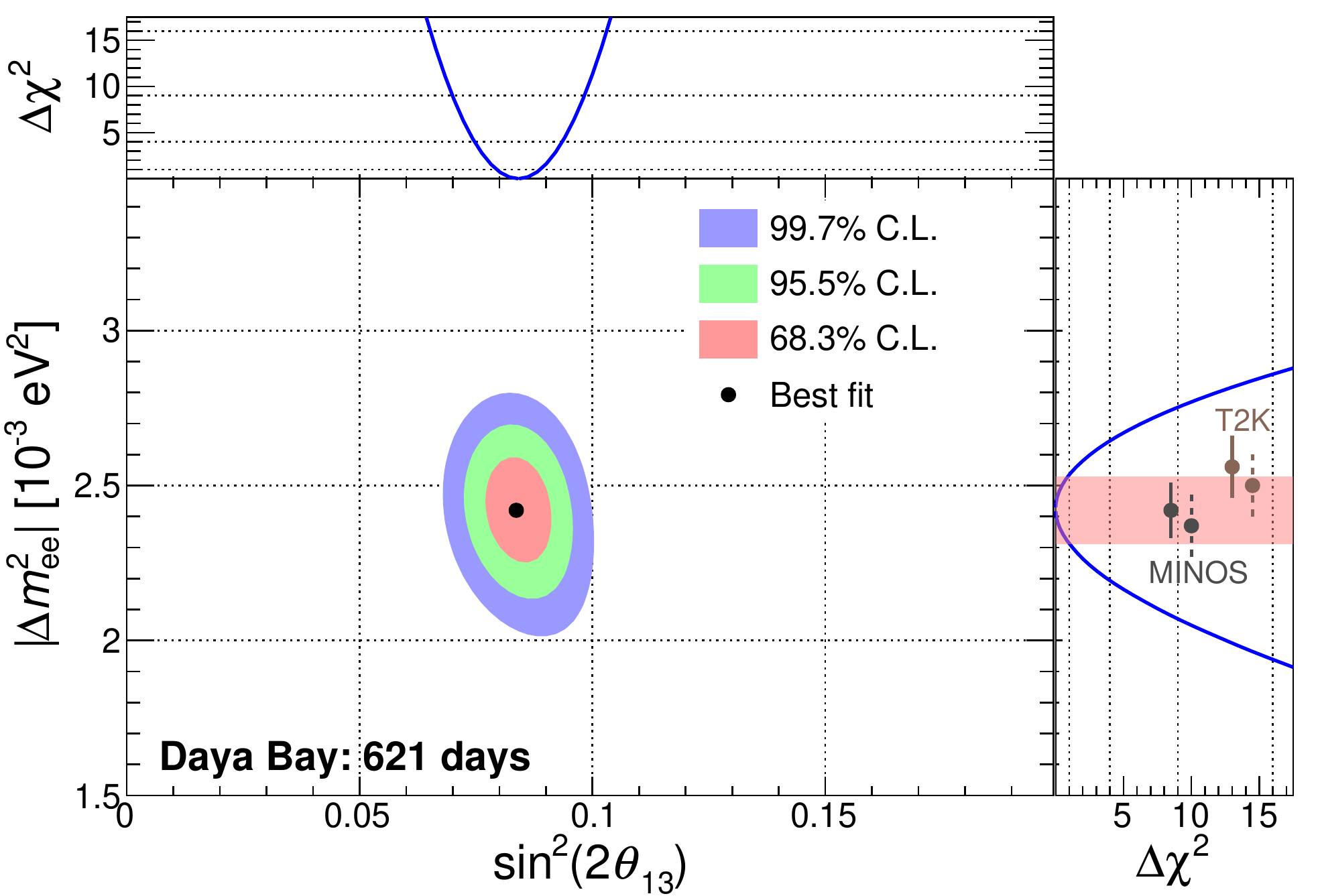}
\caption{Regions of $68.3\%$, $95.5\%$ and $99.7\%$ confidence levels in the $|\Delta m^{2}_{\mathrm{ee}}|$-$\sin^{2}2\theta_{13}$ parameter space. The adjoining panels show the dependence of $\Delta \chi^{2}$ on $\sin^{2}2\theta_{13}$ (top) and $|\Delta m^2_{\mathrm{ee}}|$ (right).
The $|\Delta m^2_{\mathrm{ee}}|$ allowed region (shaded band, $68.3\%$
C.L.) was consistent with measurements of $|\Delta m^2_{32}|$ using muon neutrino disappearance by the MINOS~\cite{Adamson:2014vgd} and T2K~\cite{Abe:2014ugx} experiments,
converted to $|\Delta m^2_{\mathrm{ee}}|$ assuming the normal (solid)
and inverted (dashed) mass hierarchy.}
\label{fig:contours}
\end{figure}

Oscillation parameters are extracted by a relative measurement of prompt energy spectra at the near and far halls. A covariance matrix including known systematic and statistical uncertainties was used in a $\chi^2$ fit. We found $\sin^22\theta_{13} = 0.084\pm0.005$ and $|\Delta m^2_{ee}|=2.42^{+0.10}_{-0.11}\times 10^{-3}$~eV$^2$ with $\chi^2$/NDF = 134.7/146. Fig.~\ref{fig:contours} shows the allowed regions for $\sin^22\theta_{13}$ and $|\Delta m^2_{ee}|$ at the 68.4\%, 95.5\%, and 99.7\% confidence levels.
It is worth noting that the current accuracy is still dominated by statistics at the far site. The $|\Delta m^2_{ee}|$ measurement is highly consistent with and of comparable precision to the muon neutrino disappearance experiments~\cite{Adamson:2014vgd,Abe:2014ugx}. Under the normal (inverted) hierarchy assumption, $|\Delta m^2_{ee}|$ yields $|\Delta m^2_{32}|=2.39^{+0.10}_{-0.11}$~eV$^2$ ($|\Delta m^2_{32}|= -2.49^{+0.10}_{-0.11}$~eV$^2$).

\subsection{Oscillation Analysis from Neutron Capture on Hydrogen}

\begin{figure}[!t]
\centering
\includegraphics[width=0.6\columnwidth]{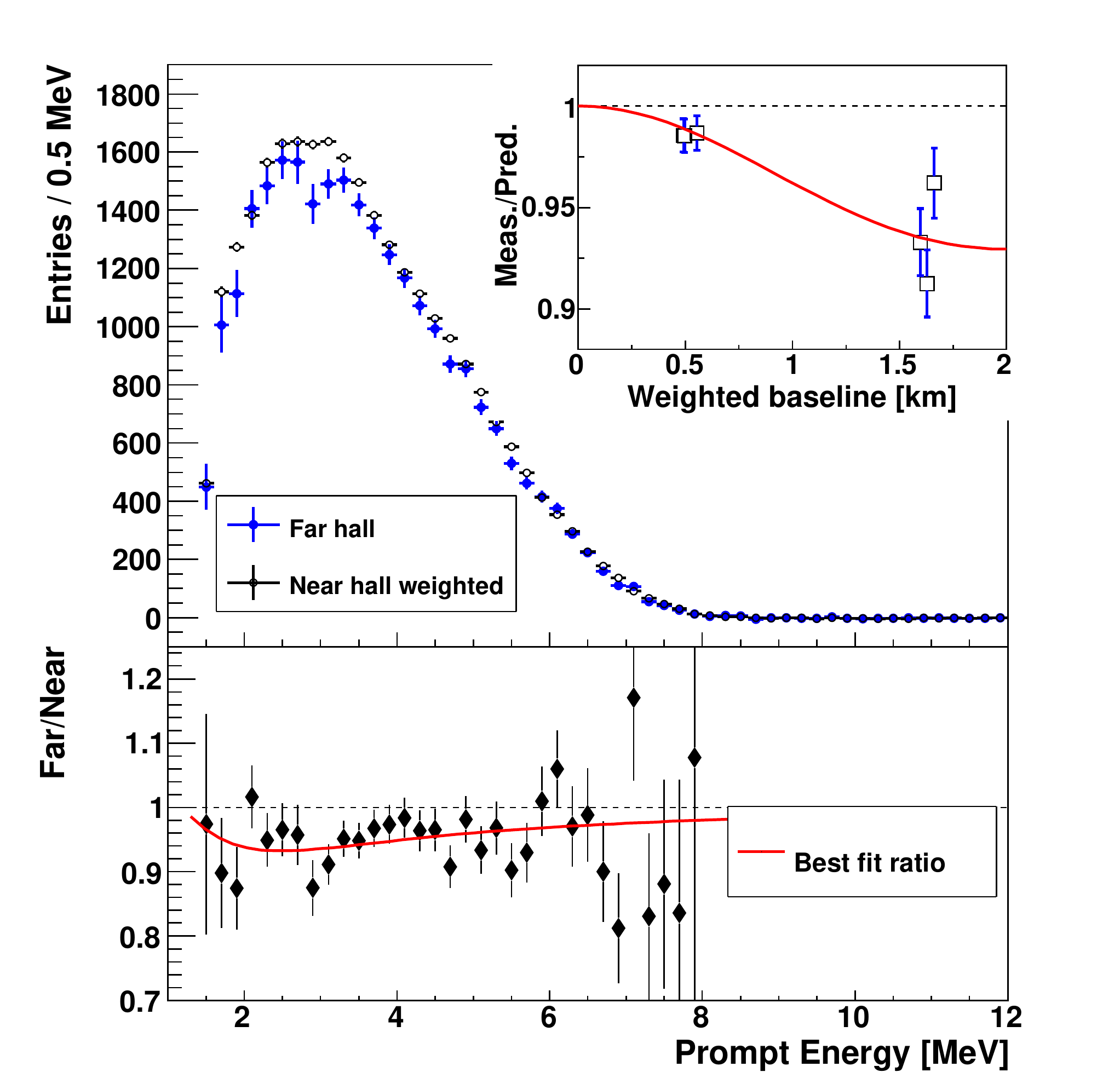}
\caption{
The observed prompt energy spectrum of the far hall ADs (blue) and
near hall ADs (open circle) weighted according to baseline.
The far-to-near ratio (solid dot) with best fit $\theta_{13}$ value is shown in the lower panel. In the inset is the ratio of the measured to the predicted rates in
each AD {\it vs.}\ baseline, in which the baselines of far ADs were shifted relative to each other by 30 ($-30$) m for clarity. }
\label{fig:NF}
\end{figure}

An independent measurement of $\sin^22\theta_{13}$ has been obtained via the detection of IBDs tagged by neutron capture on hydrogen (nH). Comparable statistics as the nGd case is achieved by the $\sim$15\% of neutron captures in the Gd-LS region and almost all of the neutron captures in the LS region. New analysis approaches have been developed to meet the
challenges associated with the higher background, longer neutron capture time ($\sim$200~$\mu$s), and a lower energy (2.2~MeV) $\gamma$ ray from neutron capture for nH IBD events. The prompt candidate is required to have $E > 1.5$~MeV, a longer coincidence
time window of 400~$\mu$s is required, and the distance between the prompt and delayed candidates is required to be $<0.5$~m. Based on the rate deficit observed at the far-site detectors for the 6-AD period, a value of $\sin^22\theta_{13}=0.083\pm0.018$ is extracted. The result is shown in Fig.~\ref{fig:NF}, which is consistent with the result obtained in the nGd analysis.

\subsection{Measurement of the Reactor Antineutrino Flux and Spectrum}
\begin{figure}[htb]
\centering
\includegraphics[width=0.45\columnwidth]{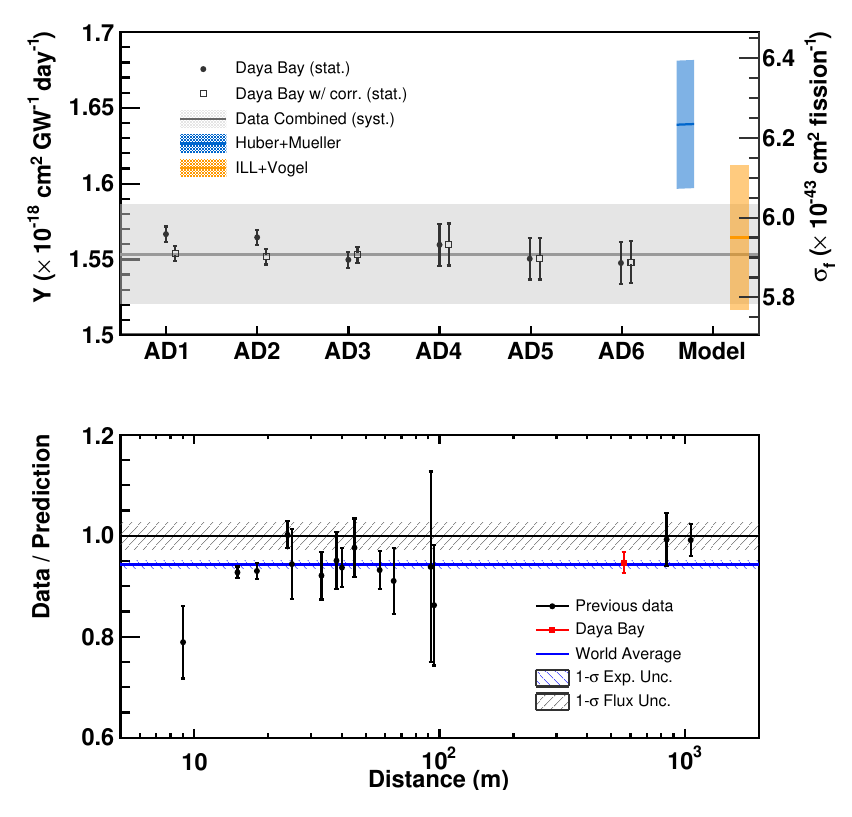}
\includegraphics[width=0.45\columnwidth]{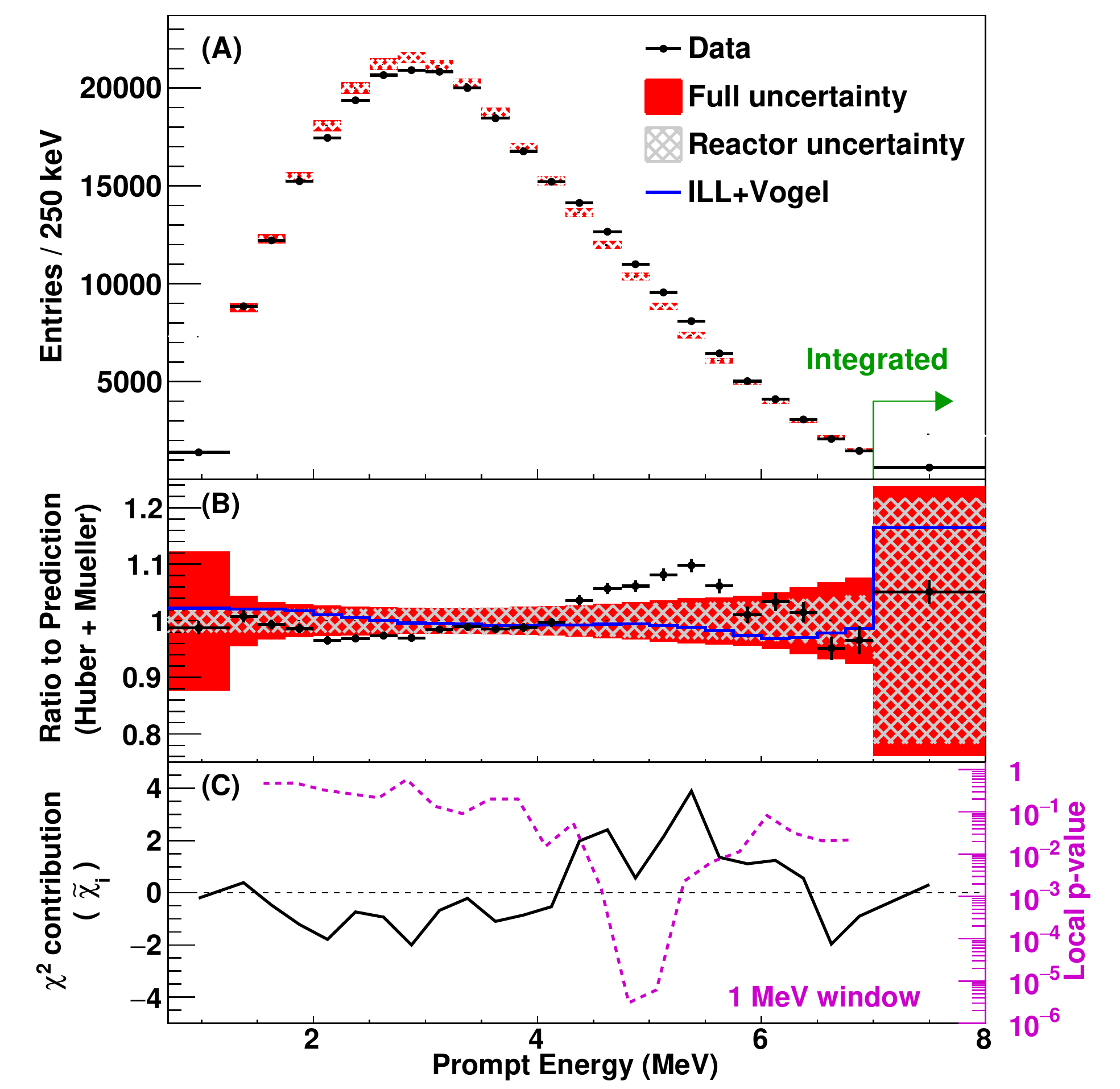}
\caption{Left: (Top panel) The measured $\bar{\nu}_e$ rate in the six ADs (closed circles), with correction of flux-weighted fission fractions at different sites (open squares). The average ( uncertainty) of three near detectors is shown in gray line (band). The rate predicted by Huber+Mueller (ILL+Vogel) model and its uncertainty are shown in blue (orange) region. (Bottom panel) The measured reactor $\bar{\nu}_e$ rate as a function of the distance from the reactor, normalized to the Huber+Mueller model. The blue shaded region represents the global average and its $1\sigma$ uncertainty. The 2.7\% model uncertainty is shown as a band around unity.
Right: (Panel A) Predicted and measured prompt energy spectra.
  The prediction is based on the Huber+Mueller model.
  The gray hatched (only in panel B) and red filled bands
  represent the square-root of diagonal elements of
  the covariance matrix ($\sqrt{(V_{ii})}$) for the reactor related and
  the full systematic uncertainties, respectively.
  The error bars on the data points represent the statistical uncertainty.
  (Panel B)
  Ratio of the measured prompt energy spectrum to the predicted spectrum
  (Huber+Mueller model).
  The blue curve shows the ratio of the prediction of ILL+Vogel model to the Huber+Mueller model.
  (Panel C) The defined $\chi^2$ distribution ($\widetilde{\chi_i}$) of
  each bin (black solid curve) and local p-values for 1 MeV energy
  windows (magenta dashed curve). See Ref.~\cite{6AD_reactor}
  for the definitions of those variables.
}
\label{fig:norm}
\end{figure}

In the complete 6-AD period, 296721 (41589) IBD candidates were detected in the near (fall) halls, and were used for a precise measurement of reactor antineutrino flux and spectrum.  The measured IBD yield, expressed as  $Y = (1.55 \pm 0.04) \times 10^{-18}$~cm$^2$/GW/day, or $\sigma_f = (5.92 \pm 0.14) \times 10^{-43}$~cm$^2$/fission, is shown in the left top plot of Fig.~\ref{fig:norm}. The measured IBD yields are consistent among all ADs. A correlated uncertainty in the detection
efficiency of 2.1\% is the dominant factor for the overall uncertainty of 2.3\%. A global fit for $R$, defined as the ratio of the Daya Bay measured rate to the model prediction, was performed to compare with the previous reactor antineutrino flux measurements following the method described in Ref.~\cite{Zhang:2013ela}. The Huber~\cite{huber} and ILL~\cite{ILL1, ILL2} models predict the reactor $\overline{\nu}_{e}$ spectra for $^{235}$U, $^{239}$Pu, and $^{241}$Pu, while the Mueller~\cite{mueller} and Vogel~\cite{vogel} models predict it for $^{238}$U. We obtained $R=0.946\pm0.022$ ($0.991\pm0.023$) for the Huber+Mueller (ILL+Vogel) model, shown in the left bottom plot of Fig.~\ref{fig:norm}.

In addition to the reactor antineutrino flux, the energy spectrum was measured and compared with the model prediction shown in the right plot of Fig.~\ref{fig:norm}. The spectral discrepancy around 5 MeV prompt energy is visible. The discrepancy is found to be time-independent and correlated with reactor power, therefore disfavoring hypotheses involving detector response and unknown backgrounds. A recent ab-initio  calculation of the antineutrino spectrum shows a similar deviation in the 4-6~MeV energy region~\cite{Dwyer:2014eka}.

\subsection{Search for a Light Sterile Neutrino}
\begin{figure}[!htb]
\centering
\includegraphics[width=0.6\columnwidth]{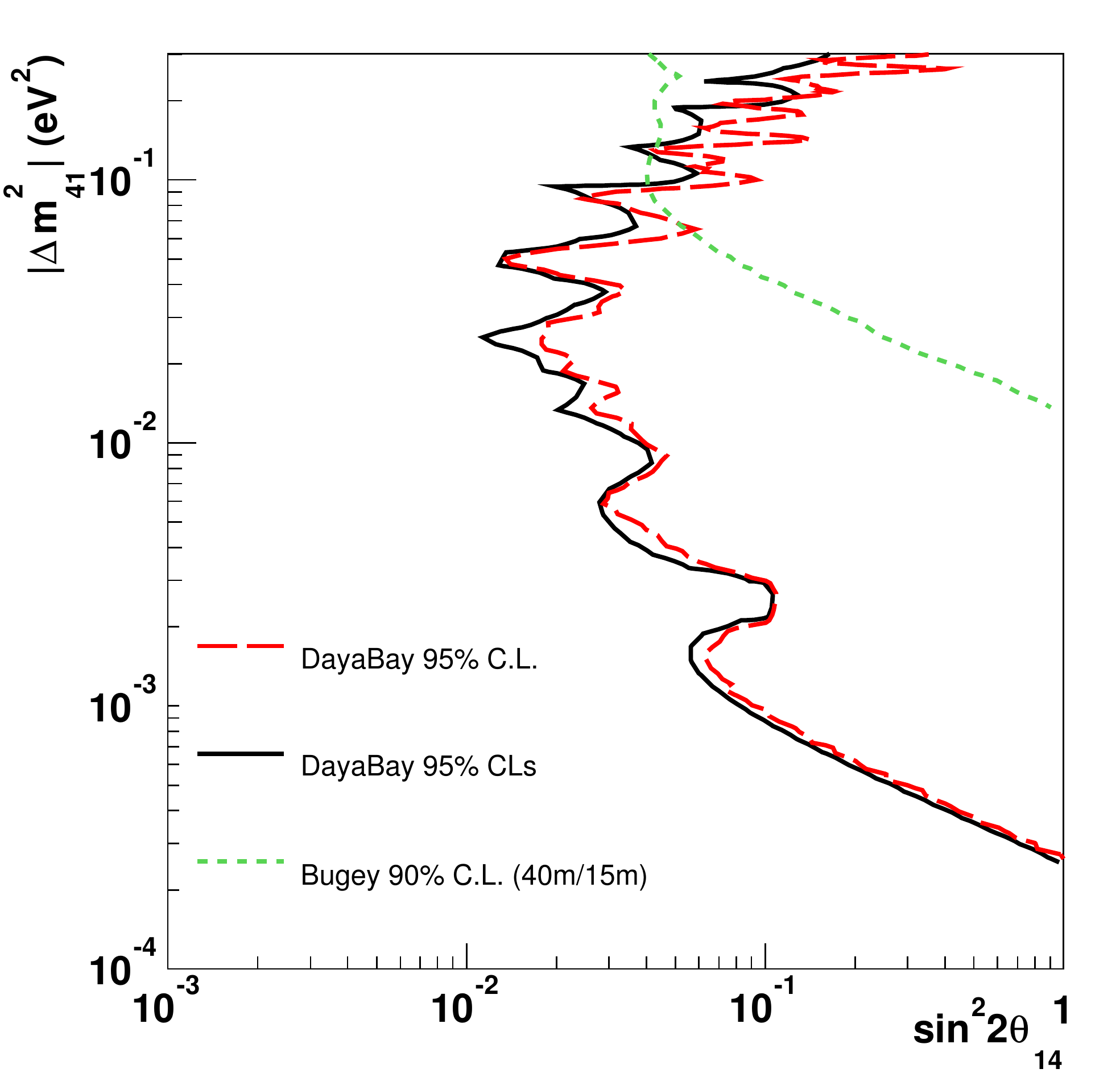}
\caption{
Exclusion contours for the neutrino oscillation parameters
  $\sin^{2}2\theta_{14}$ and $|{\Delta}m^{2}_{41}|$.
Normal mass hierarchy is assumed for both $\Delta m_{31}^{2}$ and $\Delta m_{41}^{2}$.
The red long-dashed curve represents the 95\% C.L.
exclusion contour with the Feldman-Cousins method~\cite{Feldman:1997qc}.
The black solid curve represents the 95\% ${\rm CL_{s}}$ exclusion
contour~\cite{Read:2002hq, Qian:2014nha}. The parameter-space to the right side of the contours are excluded. For comparison, Bugey's~\cite{Declais:1994su} 90\% C.L. limit on $\overline{\nu}_e$ disappearance is also shown as the green dashed curve.}
\label{fig:SpectralRatio}
\end{figure}

A search for light sterile neutrino mixing was performed in the 6-AD period. If the sterile (fourth) neutrino exists, its presence could be detected via the modification
to the three-neutrino oscillatory behavior. The experiment's unique configuration of multiple
baselines from six reactors to six ADs makes it possible to test for oscillations to a sterile neutrino in the $10^{-3}~{\rm eV}^{2} < |\Delta m_{41}^{2}| < 0.3~{\rm eV}^{2}$ range.
The relative spectral distortion due to electron antineutrino disappearance was found to be consistent with that of the three-flavor oscillation model. Fig.~\ref{fig:SpectralRatio} shows the derived limits on $\sin^22\theta_{14}$ covering the $10^{-3}~{\rm eV}^{2} \lesssim |\Delta m^{2}_{41}| \lesssim 0.1~{\rm eV}^{2}$ region, which was largely unexplored.

\section{Summary}
After the final two of eight antineutrino detectors were installed in the summer of 2012,  an additional 404 days of data was acquired with the full 8-AD configuration. The uncertainties of $\sin^22\theta_{13}$ and $|\Delta m^2_{ee}|$ were halved as a result of improvements in calibration, background estimation, as well as increased statistics. An analysis of the relative antineutrino rates and energy spectra between detectors yielded $\sin^22\theta_{13} = 0.084\pm0.005$ and $|\Delta m^2_{ee}|=2.42^{+0.10}_{-0.11}\times 10^{-3}$~eV$^2$. This is the most precise measurement of $\sin^22\theta_{13}$ to date and the most precise measurement of of $|\Delta m^2_{ee}|$ via electron antineutrino disappearance. The Daya Bay experiment is scheduled to continue data-taking with all 8 ADs until the end of 2017 at least, when the precision on both parameters is expected to reach $\sim$3\%.

Several other analysis results are presented. An independent measurement of $\sin^22\theta_{13}$  via the detection of IBDs tagged by neutron capture on hydrogen yielded $\sin^22\theta_{13}=0.083\pm0.018$, which was consistent with the nGd result. A precise measurement of reactor antineutrino flux and spectrum has been performed. A search for sterile neutrino mixing set stringent limits in the $10^{-3}~{\rm eV}^{2} < |\Delta m_{41}^{2}| < 0.3~{\rm eV}^{2}$ range, which was largely unexplored. An improvement in the reach of all of these analyses is expected as more statistics are collected and the systematic errors is reduced.

\section*{Acknowledgments}
Daya Bay is supported in part by the Ministry of Science and Technology of China, the U.S. Department of Energy, the Chinese Academy of Sciences, the CAS Center for Excellence in Particle Physics, the National Natural Science Foundation of China, the Guangdong provincial government, the Shenzhen municipal government, the China General Nuclear Power Group, Key Laboratory of Particle and Radiation Imaging (Tsinghua University), the Ministry of Education, Key Laboratory of Particle Physics and Particle Irradiation (Shandong University), the Ministry of Education, Shanghai Laboratory for Particle Physics and Cosmology, the Research Grants Council of the Hong Kong Special Administrative Region of China, the University Development Fund of The University of Hong Kong, the MOE program for Research of Excellence at National Taiwan University, National Chiao-Tung University, and NSC fund support from Taiwan, the U.S. National Science Foundation, the Alfred~P.~Sloan Foundation, the Ministry of Education, Youth, and Sports of the Czech Republic, the Joint Institute of Nuclear Research in Dubna, Russia, the CNFC-RFBR joint research program, the National Commission of Scientific and Technological Research of Chile, and the Tsinghua University Initiative Scientific Research Program. We acknowledge Yellow River Engineering Consulting Co., Ltd., and China Railway 15th Bureau Group Co., Ltd., for building the underground laboratory. We are grateful for the ongoing cooperation from the China General Nuclear Power Group and China Light and Power Company.


\begin{thebibliography}{99}
\bibitem{DYB1} F. P. An {\it et al.} (Daya Bay collaboration), Phys.\ Rev.\ Lett.\ {\bf 108}, 171803 (2012).

\bibitem{T2K} K. Abe {\it et al.} (T2K collaboration), Phys.\ Rev.\ Lett.\ {\bf 107}, 041801 (2011).

\bibitem{MINOS} P. Adamson {\it et al.} (MINOS collaboration), Phys.\ Rev.\ Lett.\ {\bf 107}, 181802 (2011).

\bibitem{DC1} Y. Abe {\it et al.} (Double Chooz collaboration), Phys.\ Rev.\ Lett.\ {\bf 108}, 131801 (2012).

\bibitem{RENO} J. K. Ahn {\it et al.} (RENO collaboration), Phys.\ Rev.\ Lett.\ {\bf 108}, 191802 (2012).

\bibitem{DYB_cdr}
  X.~Guo {\it et al.}  [Daya Bay Collaboration],
  hep-ex/0701029.

\bibitem{DYB_nima}
  F.~P.~An {\it et al.}  [Daya Bay Collaboration],
  Nucl.\ Instrum.\ Meth.\ A {\bf 685} (2012) 78

\bibitem{DYB_muon}
  F.~P.~An {\it et al.}  [Daya Bay Collaboration],
  Nucl.\ Instrum.\ Meth.\ A {\bf 773}, 8 (2015)

\bibitem{DYB_cpc}
  F.~P.~An {\it et al.}  [Daya Bay Collaboration],
  Chin.\ Phys.\ C {\bf 37}, 011001 (2013)


\bibitem{DYB_6AD_shape}
  F.~P.~An {\it et al.}  [Daya Bay Collaboration],
  Phys.\ Rev.\ Lett.\  {\bf 112}, 061801 (2014)

\bibitem{DYB_8AD}
  F.~P.~An {\it et al.}  [Daya Bay Collaboration],
  arXiv:1505.03456 [hep-ex].


\bibitem{DYB_nH}
  F.~P.~An {\it et al.}  [Daya Bay Collaboration],
  Phys.\ Rev.\ D {\bf 90}, no. 7, 071101 (2014)

\bibitem{6AD_reactor}
W. L. Zhong (Daya Bay Collaboration), Presentation given at ICHEP2014.

\bibitem{DYB_sterile}
  F.~P.~An {\it et al.}  [Daya Bay Collaboration],
  Phys.\ Rev.\ Lett.\  {\bf 113}, 141802 (2014)


\bibitem{Adamson:2014vgd}
  P.~Adamson {\it et al.}  [MINOS Collaboration],
  Phys.\ Rev.\ Lett.\  {\bf 112}, 191801 (2014)

\bibitem{Abe:2014ugx}
  K.~Abe {\it et al.}  [T2K Collaboration],
  Phys.\ Rev.\ Lett.\  {\bf 112}, no. 18, 181801 (2014)

\bibitem{Zhang:2013ela}
  C.~Zhang, X.~Qian and P.~Vogel,
  Phys.\ Rev.\ D {\bf 87}, no. 7, 073018 (2013)

\bibitem{huber}
  P.~Huber,
  Phys.\ Rev.\ C {\bf 84}, 024617 (2011)
  [Phys.\ Rev.\ C {\bf 85}, 029901 (2012)]

\bibitem{ILL1}
  K.~Schreckenbach, G.~Colvin, W.~Gelletly and F.~Von Feilitzsch,
  Phys.\ Lett.\ B {\bf 160}, 325 (1985).

\bibitem{ILL2}
  A.~A.~Hahn, K.~Schreckenbach, G.~Colvin, B.~Krusche, W.~Gelletly and F.~Von Feilitzsch,
  Phys.\ Lett.\ B {\bf 218}, 365 (1989).

\bibitem{mueller}
  T.~A.~Mueller, D.~Lhuillier, M.~Fallot, A.~Letourneau, S.~Cormon, M.~Fechner, L.~Giot and T.~Lasserre {\it et al.},
  Phys.\ Rev.\ C {\bf 83}, 054615 (2011)

\bibitem{vogel}
  P.~Vogel, G.~K.~Schenter, F.~M.~Mann and R.~E.~Schenter,
  Phys.\ Rev.\ C {\bf 24}, 1543 (1981).

\bibitem{Dwyer:2014eka}
  D.~A.~Dwyer and T.~J.~Langford,
  Phys.\ Rev.\ Lett.\  {\bf 114}, no. 1, 012502 (2015)

\bibitem{Feldman:1997qc}
  G.~J.~Feldman and R.~D.~Cousins,
  Phys.\ Rev.\ D {\bf 57}, 3873 (1998)

\bibitem{Read:2002hq}
  A.~L.~Read,
  J.\ Phys.\ G {\bf 28}, 2693 (2002).
  
\bibitem{Qian:2014nha}
  X.~Qian, A.~Tan, J.~J.~Ling, Y.~Nakajima and C.~Zhang,
  arXiv:1407.5052 [hep-ex].

\bibitem{Declais:1994su}
  Y.~Declais, J.~Favier, A.~Metref, H.~Pessard, B.~Achkar, M.~Avenier, G.~Bagieu and R.~Brissot {\it et al.},
  Nucl.\ Phys.\ B {\bf 434}, 503 (1995).


\end{thebibliography}
\end{document}